\documentclass[3p, review]{elsarticle}

\usepackage{amssymb}
\usepackage{gensymb}
\usepackage{ tipa }
\usepackage{multirow}

\usepackage{lineno}
\modulolinenumbers[5]
\usepackage{amsmath, color}
\usepackage{multirow, booktabs}
\usepackage{diagbox}
\usepackage{setspace, hyperref}
\usepackage{kotex}
\biboptions{sort&compress}

\journal{}

\begin{document}

\begin{frontmatter}



\title{Measurement of effective thermal conductivity of LaNi$_5$ powder packed bed}

\author[1,2]{Dong-min Kim}
\author[1]{Joong Bae Kim}
\author[1,2]{Jungchul Lee}
\author[1,2]{Bong Jae Lee\corref{cor}}

\cortext[cor]{Corresponding author}
\ead{bongjae.lee@kaist.ac.kr}

\address[1]{Department of Mechanical Engineering, Korea Advanced Institute of Science and Technology, Daejeon 34141, South Korea}
\address[2]{Center for Extreme Thermal Physics and Manufacturing, Korea Advanced Institute of Science and Technology, Daejeon 34141, South Korea}

\begin{abstract}
Effective thermal conductivity of LaNi$_5$ powder packed bed was analyzed with customized guarded hot-plate (GHP) apparatus. Here, GHP was designed for precise measurement of effective thermal conductivity of metal-hydride powders even with small sample amounts (2.12$\times$10$^4$ mm$^3$). Dimensions of sample container and apparatus were determined through two-dimensional (2-D) steady-state heat conduction analysis. Calibration experiment and uncertainty analysis were conducted to validate the accuracy of the GHP. Based on the measurements of the residual thermal conductivity of the LaNi$_5$ packed bed, effect of particle size on contact factor of LaNi$_5$ packed bed was estimated. By applying the Yagi and Kunii (YK) model to the effective thermal conductivity of LaNi$_5$ packed bed, effect of contact factor and gas thermal conductivity on characteristic length of gas film were newly analyzed. Factors of YK model were modified in present work and validated through comparison with experimental data from previous literature.  
\end{abstract}

\begin{keyword}
Metal-hydride powder, Effective thermal conductivity, Residual thermal conductivity, Guarded hot-plate apparatus

\end{keyword}

\end{frontmatter}


\section{Introduction}
Hydrogen energy has been highlighted for its eco-friendly characteristics and wide transformability. Hydrogen can play a role as an energy carrier between primary energy source and electrical power, so transportation and storage methods have been thoroughly studied in this area. Among various storage methods, chemical storage methods based on metal-hydrides have been actively studied for the past several decades due to their high energy density and safety \cite{zuttel2003materials, lototskyy2017use}. Since hydriding-dehydriding cycles of metal-hydrides involves exothermic and endothermic reactions, the temperature of a storage system limits the reaction rate of hydrogenation or dehydrognation \cite{gopal1995studies}. That is, effective heat transfer between metal-hydrides induces efficient hydrogenation reactions \cite{zhang2005review, afzal2017heat}. However, as metal-hydride alloy split into fine powders by mechanical constraints generated by lattice volume expansion during hydriding-dehydriding cycles, the effective thermal conductivity of the metal-hydride packed bed decreases drastically \cite{hahne1998thermal}. Therefore, various thermal designs for enhancing the heat transfer performance of storage systems have been developed over the past several decades \cite{yang2010high,keshari2017numerical,chibani2020heat}. Since thermal designs involving metal fins or foams take up internal space of the storage system and eventually reduce the system energy density, accurate measurement of the effective thermal conductivity of metal-hydrides is crucial.

Measurement methods for thermal conductivity of metal-hydrides can be roughly categorized into transient and steady-state methods. The transient method obtains the thermal conductivity by analyzing temperature profiles of samples while various types of heater release sudden heat pulses onto those samples \cite{ishido1982thermal, pons1993measurement, kallweit1994effective, hahne1998thermal, dedrick2005thermal, christopher2006application, merckx2012simplified}. Hahne and Kallweit \cite{hahne1998thermal} used the transient hot-wire method to measure the effective thermal conductivity of HWT5800 and LaNi$_5$. Ishido \textit{et al.}\ \cite{ishido1982thermal} also used the hot-wire method to measure the thermal conductivity of sodium alanate. Although the transient method makes it easy to measure the thermal conductivity, the hot-wire method is not appropriate for effective thermal conductivity measurement of a packed bed. The wire has insufficient contact with the metal-hydride powder packed bed, which causes measurement inaccuracy \cite{flynn1996design}. In fact, experimental studies using the transient method for effective thermal conductivity measurement suffered from relatively large uncertainty of over 10\% \cite{kapischke1998measurement, hahne1998thermal, kapischke1994measurement}. On the other hand, the steady-state method determines thermal conductivity by analyzing temperature profile based on Fourier's law or the Laplace equation \cite{suda1980experimental,kempf1986measurement, da1990theoretical, lloyd1998thermal, kumar2011measurement}. Kempf and Martin \cite{kempf1986measurement} used the steady-state method to measure the effective thermal conductivity of TiFe$_{0.85}$Mn$_{0.15}$. Kumar \textit{et al.}\ \cite{kumar2011measurement} used the steady-state method with axial heat transfer and the comparative method to measure the effective thermal conductivity of MmNi$_{4.5}$Al$_{0.5}$. In their work, guard heater and insulation material were used to induce an axial heat transfer condition, and thermocouples were used to read the temperature profile of the packed bed. In general, the steady-state method is more accurate and reliable for measurement of metal-hydride thermal conductivity; however, it usually requires a large quantity of the sample (over 10$^5$ mm$^3$) \cite{suissa1984experimental,kumar2011measurement,kempf1986measurement}, which is not appropriate for metal-hydrides under development \cite{cho2016graphene}. In addition, loss of accuracy can occur due to the inhomogeneous temperature profile of the packed bed resulting from the thermocouples inserted into the sample \cite{suissa1984experimental,suda1980experimental,kempf1986measurement}. 

Based on the measured effective thermal conductivity of the metal-hydride packed bed, modeling of the effective thermal conductivity has been widely studied for hydriding materials. A theoretical model of heat transfer is crucial for understanding heat and mass transfer in a hydrogen storage system. During the past several decades, variations of the Zehner, Bauer and Schl\"under model \cite{zehner1970warmeleitfahigkeit} and the Yagi and Kunii model \cite{yagi1957studies} were generally used for modeling of the effective thermal conductivity of powder packed beds. The Yagi and Kunii (YK) model treats the porous medium as packed spheres of the same diameter and material, while the Zehner, Bauer and Schl\"under (ZS) model can be applied to non-spherical models or mixtures of grains for different sizes and materials. The ZS model is potentially more attractive for modeling the effective thermal conductivity, but this model includes many factors that can only be achieved by experiment, which causes complexity in the fitting process of the thermal conductivity in diverse working conditions. Kallweit and Hahne \cite{kallweit1994effective} used the ZS model to calculate the thermal conductivity of LaNi$_5$ packed bed, and for this, contact area fractions considering existing literature values were used. On the other hand, Pons and Dantzer \cite{pons1994determination} used an assumed contact factor to evaluate the particle diameter and bulk thermal conductivity of an LaNi$_5$ packed bed. Both studies lacked consideration of the heat transfer properties of the LaNi$_5$ particles depend on the particle size, contact factor and thermal conductivity of the gas. 

In this work, the effective thermal conductivity of LaNi$_5$ packed bed was precisely measured with a customized guarded hot-plate (GHP) apparatus that can be working with limited sample quantity (about $2.12\times10^4$ mm$^3$). Intentionally, no thermocouples were present to interrupt the temperature profile of the packed bed. Calibration and uncertainty analysis were performed, and the accuracy of the apparatus was verified by measuring the thermal conductivity of the reference materials. With simple and accurate measurements of the effective thermal conductivity obtained by customized GHP, more quantitative analysis of the heat transfer performance of metal-hydrides became possible. In contrast to experimental studies using the YK model for metal-hydrides \cite{suissa1984experimental, kumar2011measurement}, which referred to empirical estimates which solely depend on porosity \cite{yagi1957studies}, we modified the factors for LaNi$_5$ in the YK model by solid measurement of effective thermal conductivity of the LaNi$_5$ powder packed bed. Furthermore, effects of particle size, contact factor, and gas thermal conductivity on the heat transfer characteristics of the LaNi$_5$ packed bed were newly analyzed.

\section{Experimental setup and method}
The guarded hot-plate (GHP) method is a steady-state method that uses axial heat flux to obtain thermal conductivity \cite{kobari2015development}. The GHP apparatus consists of a hot-surface and cold-surface assemblies. Specimen with sample container is placed between the hot-surface and cold-surface assemblies. The hot-surface assembly consists of a guard plate and main plate, and should be positioned above the specimen to prevent natural convection \cite{kathare2008natural}. The main plate is surrounded by the guard plate, which minimizes the heat loss of the main plate. Both hot-surface and cold-surface assemblies form isothermal boundary conditions at areas in contact with the specimen. Dissipated heat from the main plate forms a one-dimensional (1-D) heat conduction condition through the specimen. The thermal conductivity ($k_s$) of the specimen can thus be calculated by Fourier's law:
\begin{equation} \label{Eq:1}
	k_{s}=\frac{Q_{s}d_{sp}}{(T_{m}-T_{c})A_{m}}
\end{equation}
where $d_{sp}$ is the thickness of the specimen, $A_{m}$ is the surface area of the main plate, $Q_{s}$ is the heat transfer rate from the main heater to the specimen, and $T_m$ and $T_c$ are the temperatures of the main plate and cold-surface assemblies, respectively. Thus, heat loss from the main plate is one of the most important factors affecting the accuracy of the GHP apparatus, which mainly consists of heat loss through the gap between the main plate and the guard plate (i.e., $Q_{gap}$), heat conduction through lead wires (i.e., $Q_{lead}$), and heat conduction through the area in contact with the guard plate (i.e., $Q_{cont}$). 

\begin{figure}[!t]
	\centering
	\includegraphics[width=0.6\columnwidth]{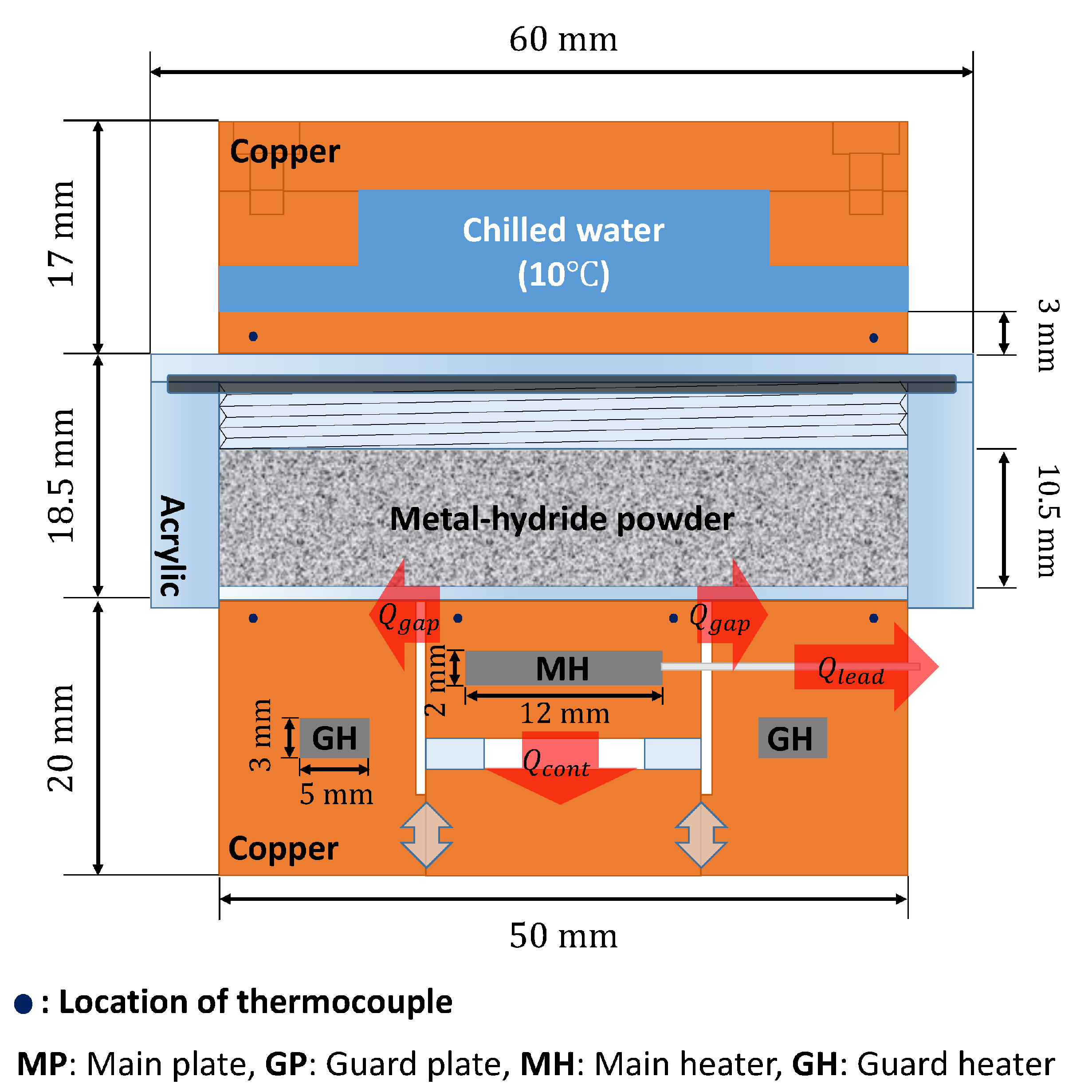}
	\caption{Schematic of the customized GHP. The hot-surface assembly consists of main plate and guard plate. Heat loss from the main plate can occur via: (1) conduction and radiation through gap between main plate and guard plate (i.e., $Q_{gap}$); (2) conduction through lead wire (i.e., $Q_{lead}$); and (3) conduction through contact area with guard plate (i.e., $Q_{cont}$).}
	\label{Fig:1}
\end{figure}

Most commercial GHP apparatuses requires specimen size to be 20 to 60 cm across and 2 to 20 cm thick \cite{flynn1996design}, so as to enhance the 1-D heat conduction condition on the main plate. However, apparatuses with large dimensions have a disadvantage in the measurement of the metal-hydrides under development, which are very limited in quantity, making it inefficient to supply sufficient amounts of test specimens \cite{cho2016graphene}. Thus, we designed a sub-miniature guarded hot-plate for easy and precise measurement of the effective thermal conductivity of metal-hydride powders (refer to Fig. \ref{Fig:1}). Since contribution of heat loss to the total amount of heat from main plate increases as apparatus become smaller, geometry of the hot-surface assembly (i.e., diameter of the main plate, gap width) and the dimensions of the sample container were carefully considered to minimize the heat loss from the main plate.

The outer diameter of the GHP was fixed at 50 mm. Material for the hot-surface  and cold-surface assemblies was copper (with $k=400$ W/m$\cdot$K \cite{kaye1993tables}), which helps the apparatus to reach the steady state quickly and increases the stability. The material for the sample container was chosen as acrylic with $k= 0.150$ W/m$\cdot$K, which lies in the range of thermal conductivity of metal-hydrides packed bed \cite{kumar2011measurement}. The thermal conductivity of acrylic was measured by using a commercial instrument (Netzsch LFA 457). Here, the similarity between the thermal conductivity of metal-hydrides and the sample container is key to guarantee the 1-D conduction condition in the packed bed. As depicted in Fig.\ \ref{Fig:1}, the height of both the hot-surface and cold-surface assemblies was 20 mm, and the height of the main plate was set to be 10 mm. The main heater and guard heater were determined to use commercial heaters (WATLOW ULTRAMIC CER-1-02-0001 for guard heater and CER-1-01-00540 for main heater). The gap width between main plate and guard plate was set to be 0.75 mm considering manufacturability.  

To determine the diameter of the main plate, the axisymmetric two-dimensional (2-D) steady-state heat conduction analysis has been performed to obtain temperature distribution at the top surface of the hot-surface assembly with respect to the diameter of the main plate in a range of $18\sim24$ mm. In the calculation, diameter of the sample container is set to be 60 mm, and thermal conductivity of the sample was assumed as 0.15 W/m$\cdot$K. When the diameter of the main plate is smaller than 20 mm, the calculation revealed that the top surface temperature deviates below 5 mK from the average temperature of the hot surface. Also, difference of the mean temperature between the main plate and the guard plate was minimized  when $d_{mp}=20$ mm. Therefore, diameter of the main plate was determined to be 20 mm to minimize the heat transfer between the main plate and the guard plate. It is worthwhile to mention that when the powder has a thermal conductivity of 0.02 W/m$\cdot$K, the sample case with a large diameter (i.e., $d_c=60$ mm) improves the straightness of the heat flux by 2\% compared to using a sample case with a small diameter (i.e., $d_c=50$ mm). Thus, the outer diameter of the sample container is 10 mm larger than the outer diameter of the guard plate in this work. 



To minimize the heat loss from the main plate to the environment, the customized GHP was placed in a vacuum chamber whose pressure can be maintained under 10 mtorr. The temperature of cold-surface assembly was maintained at 10$\celsius$ by refrigerating bath circulator (Lab companion, RW3-0525G), and power supply (Agilent E3634A) was used to supply voltage to the main heater and guard heater. The heat generation rate of the heater was controlled by PID feedback so that the temperature read by the thermocouples located on the hot-surface assembly was maintained at 60$\celsius$. The thermocouples were inserted 2 mm below the surface and 3 mm deep from the outer diameter of the hot-surface and cold-surface assemblies (refer to Fig.\ \ref{Fig:1}). To increase the accuracy of measurements, no thermocouples were inserted into the sample container. The sample container was clamped between the hot-surface and cold-surface assemblies. For stable contact between main plate and the sample container, The main plate was designed to move freely up and down (refer to Fig.\ \ref{Fig:1}). After the sample container was clamped, the main plate was fixed with a set screw. Thermal paste (CANS HSC611) was conducted to every contact area.

\begin{figure}[!t]
	\centering
	\includegraphics[width=0.6\columnwidth]{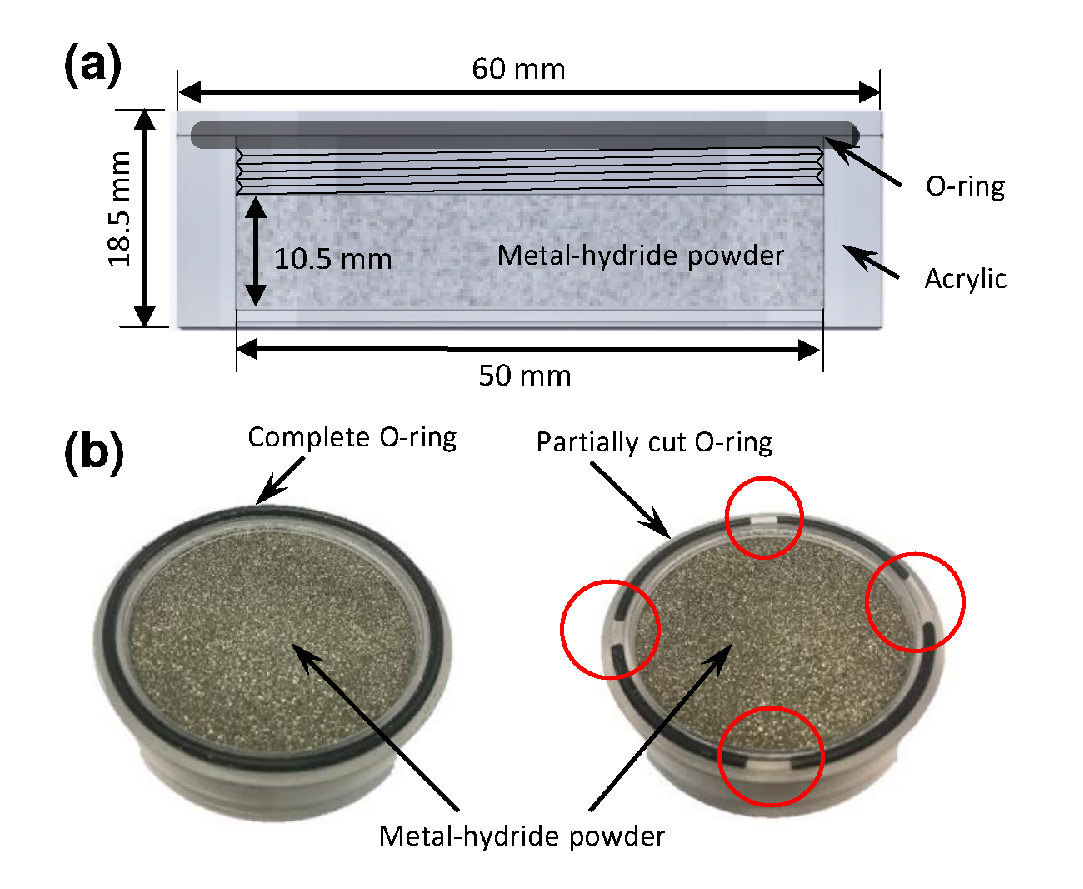}
	\caption{(a) Schematic of the sample container. Metal-hydride powers of $2.12\times 10^4$ mm$^3$ can be loaded in the container. (b) Pictures showing two sample containers for different experimental conditions. One with the complete O-ring is for measurements at atmospheric pressure, and the other with the partially cut O-ring is for measurements at vacuum ($\sim10$ mtorr).}
	\label{Fig:2}
\end{figure}

Figure \ref{Fig:2}(a) illustrates the sample container used in this study. The sub-miniature GHP apparatus developed here takes only $2.12\times10^4$ mm$^3$ of metal-hydrides, which is almost an order of magnitude less amount compared to previous studies \cite{suissa1984experimental, kempf1986measurement, kumar2011measurement}. Since the GHP is placed in the vacuum chamber, we designed two sample containers, as shown in Fig.\ \ref{Fig:2}(b). By using an air-tight (i.e., complete) O-ring, the effective thermal conductivity of the LaNi$_5$ packed bed can be measured in various gas environment at atmospheric pressure. Alternatively, if a partially cut O-ring is used, one can measure the effective thermal conductivity of LaNi$_5$ packed bed in vacuum condition (i.e., residual thermal conductivity) because the pressure inside the sample container will be the same as that in the vacuum chamber. 

\begin{table}[!b]
	\caption{Calibration experiments with acrylic ($k=0.150$ W/m$\cdot$K) when temperature difference between hot-surface and cold-surface assemblies is maintained at 50\celsius.}
	\label{Tab:2}
	\small
	\begin{center}
		\begin{tabular} {c c c c c c| c c c}
		 \hline\hline & $k_s$ & $A_m$    & $d_{sp}$  & $T_{m}-T_{g}$    & $Q_s$ & $T_m-T_c$  & $Q_m$  & $Q_{loss}$  \\
		  &(W/m$\cdot$K) & (m$^2) $   &(m) &(K)  &(W)&(K)      &(W)   &(W)\\
		 \hline
1&0.150 &0.0003 & 0.105     & $\pm 0.02 $  &   0.123 &48.14       &   0.201      &  0.078     \\
2&0.150&0.0003 & 0.105     & $\pm 0.02 $ & 0.123 & 48.11          & 0.201        & 0.078     \\
3&0.150&0.0003 & 0.105     & $\pm 0.02 $ & 0.123 & 48.11      & 0.207            & 0.085     \\
4&0.150&0.0003 & 0.105     & $\pm 0.02 $  & 0.123& 48.10       & 0.207           & 0.084     \\
5&0.150&0.0003 & 0.105     & $\pm 0.02 $  & 0.123 & 48.10      & 0.205           & 0.082     \\
\hline
  & &   &           &   & Avg.      & 48.11     & 0.204        & 0.082    \\
&  &  &           &     & Stdev.    & 0.02 & 0.003  & 0.003 \\
\hline\hline
		\end{tabular}
	\end{center}
\end{table} 

\begin{figure}[!t]
	\centering
	\includegraphics[width=0.7\columnwidth]{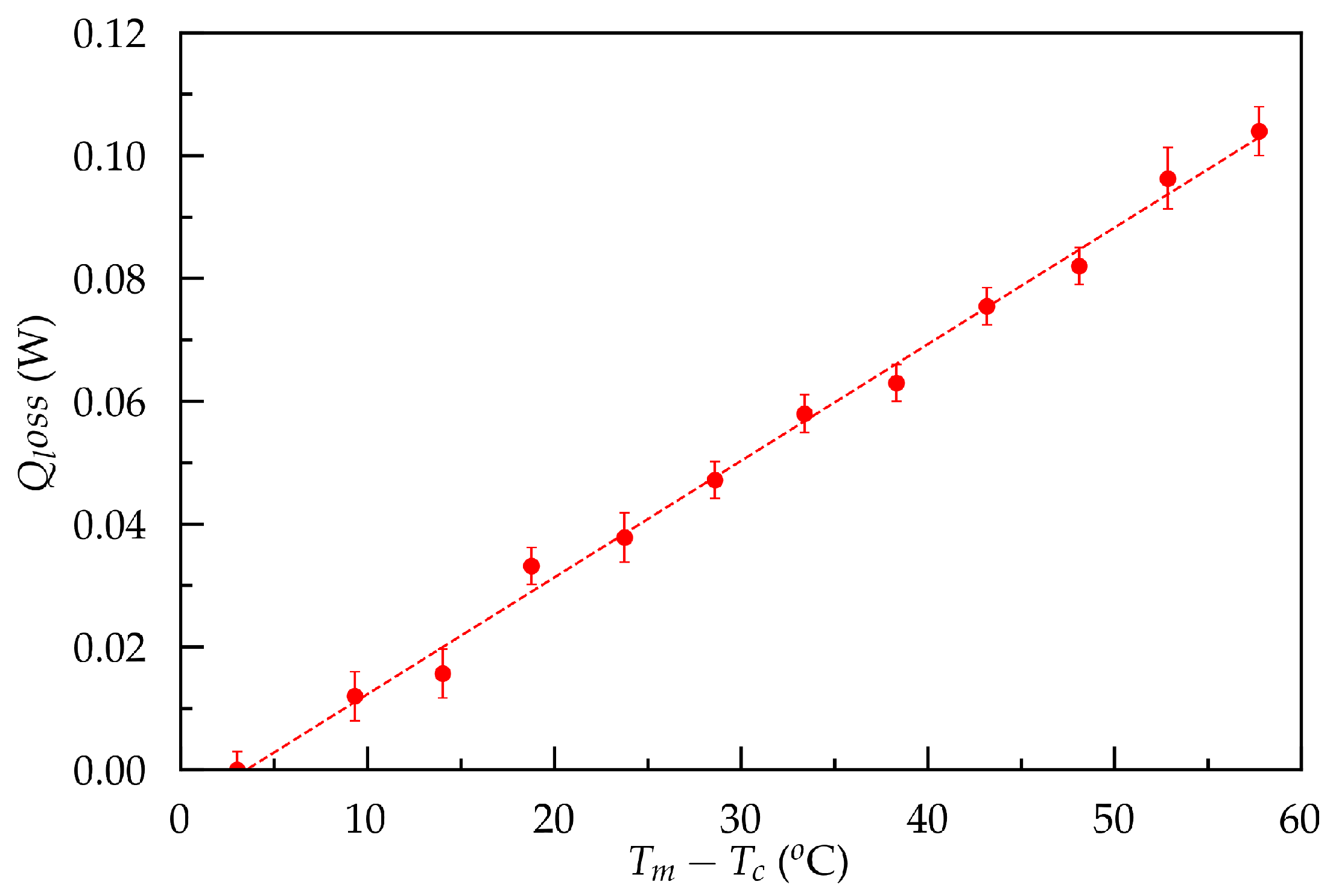}
	\caption{Heat loss ($Q_{loss}$) from the main plate respect to temperature difference between the main plate and the cold-surface assembly. Linear relationship implies that the main plate is connected to the one heat reservoir with constant thermal resistance. }
	\label{Fig:3}
\end{figure}

Calibration experiment was conducted to estimate the heat loss of the main plate. Heat loss should be carefully estimated for precise measurement of thermal conductivity. The GHP apparatus has been calibrated with acrylic whose thermal resistance is known from commercial apparatus (Netzsch LFA 457). The heat loss of the main plate can be obtained by subtracting the theoretical amount of heat passing through the specimen from the measured power of the main heater. Table \ref{Tab:2} lists the temperature difference between main plate and guard plate, measured power of the heater, heat passing through specimen and resulting heat loss of the main plate while temperature difference between main plate and cold-surface assembly is 50\celsius. The temperature difference between guard plate and main plate was maintained at 0 K with uncertainty of 0.02 K. Through five sets of measurements, the heat loss of the main plate was obtained with an uncertainty of 0.003 W. In a same way, the heat loss of the main plate with respect to the temperature difference between the main plate and the cold-surface assembly was measured. As shown in Fig.\ \ref{Fig:3}, the heat loss linearly depends on the main plate temperature, which implies that the main plate is connected to the heat reservoir at a constant temperature with a constant heat resistance. In order to satisfy this condition, the heat path from the main plate to the cold-surface assembly should be constant regardless of the temperature of the main plate, that is, the 1D heat conduction condition was well maintained during experiment. Hereafter, we use the actual heat transfer from the main heater to the sample as $Q_{s}=Q_{m}-Q_{loss}$, where $Q_m$ is the input power to the main heater.

Uncertainty analysis was conducted for the apparatus according to Refs.\ \cite{taylor1994guidelines, jcgm2008100}. According to Eq.\ \eqref{Eq:1}, the effective thermal conductivity of metal-hydrides is determined by the temperature difference of the hot-surface and cold-surface assemblies ($T_m-T_c$), power from the main heater to the sample, and the heat loss of the main plate (i.e., $Q_{s}=Q_m-Q_{loss}$). Since these three variables are independent, the combined uncertainty $u_c$ can be expressed as:  
\begin{equation} \label{Eq:2}
	u_c=\sqrt{\sum^n_i(c_i^2 u_i^2)}
\end{equation}
where $c_i$ is the sensitivity coefficient of the variables given as $c_i=\frac{\partial k_s}{\partial x_i}$ with $x_i$ being the variable (i.e., $Q_m$, $Q_{loss}$, or $\Delta T$), and $u_i$ is the uncertainty of each variables, which can be estimated from the root sum square of Type A and Type B uncertainties. Type A evaluation of standard uncertainty is based on the statistical method of treating data. On the other hand, Type B evaluation is based on the standard deviation of the assumed system distribution, which also called a systematic effect. By counting the deviation of the measured data (i.e., $Q_s$ or $\Delta T$) from its mean value after it reached the steady state, both data show that measuring systems can be safely assumed as Gaussian distributions. We set the interval as 95\% of the input data, and the standard deviation of the Gaussian distribution is half the width of the interval. 
\begin{table}[!t]
	\caption{Uncertainty analysis of the thermal conductivity measurement of PEEK ($k=0.253$ W/m$\cdot$K)}
	\label{Tab:3}
	\small
	\begin{center}
		\begin{tabular} {c c c | c c | c c | c  c | c} 
		\hline\hline
		                 & \multicolumn{2}{c}{$\Delta T$} & \multicolumn{2}{c}{$Q_m$} &\multicolumn{2}{c}{$Q_{loss}$ } & $u_c$ & $k_s$ & error \\
		                 & \multicolumn{2}{c}{($u_i=0.02$ K)}& \multicolumn{2}{c}{($u_i=0.003$ W)} & \multicolumn{2}{c}{($u_i=0.003$ W)} & (\%) & (W/m$\cdot$K) & (\%)  \\
		\hline
                    &   $x_i$    & $c_i$      & $x_i$    &    $c_i$   & $x_i$      & $c_i$ & & &     \\
1 & 47.94 & 0.013 & 0.243 & 3.77 & 0.082 & 3.77 & 6.13 & 0.263 & 3.48\\
2 & 47.95 & 0.012 & 0.238 & 3.47 & 0.082 & 3.47 & 6.08 & 0.245 & 2.69\\
3 & 47.91 & 0.012 & 0.239 & 3.54 & 0.082 & 3.54 & 6.08 & 0.249 & 1.07\\
4 & 47.94 & 0.012 & 0.240 & 3.59 & 0.082 & 3.59 & 6.10 & 0.252 & 0.73\\
5 & 47.88 & 0.011 & 0.237 & 3.44 & 0.082 & 3.44 & 6.08 & 0.242 & 4.36\\
   \hline\hline    
		\end{tabular}
	\end{center}
\end{table} 

As a validation of the customized GHP apparatus, uncertainty analysis was performed on thermal conductivity measurements of PEEK ($k= 0.253$ W/m$\cdot$K). Thermal conductivity of the PEEK was measured in advance by laser flash method (Netzh LFA 457). Values and sensitivity coefficients of variables for each measurement are listed on Table \ref{Tab:3}. The mean value of the thermal conductivity for five set of measurements was obtained as 0.250 W/m$\cdot$K, which deviate 1.1\% from the reference value. To define the uncertainty of the mean value, we set the combined uncertainty ($u_c$) and the standard deviation of the five set of measurements as variables in Eq. (\ref{Eq:2}). The combined uncertainty of one measurement was calculated as $\sim 6$\% of thermal conductivity of the PEEK. Meanwhile, we have achieved 3\% of standard deviation through five set of measurements. Since the average value is obtained by five measurements, the sensitivity coefficient of combined uncertainty is 0.2. Thus, uncertainty of the mean value was achieved as 4.2\% of reference value, which is almost $3 \sim 4$ times smaller uncertainty compared to previous studies \cite{suissa1984experimental,kempf1986measurement,kapischke1998measurement}.

\section{Results and discussion}
Considering that the effective thermal conductivity of the LaNi$_5$ powder packed bed can vary with its porosity and contact factors, LaNi$_5$ powder (Horizon Fuel Cell Technologies Co. Ltd.) was first classified according to its particle size using a sieve. Table \ref{Tab:4} list the classified LaNi$_5$ powder sample . For the thermal conductivity measurement, the LaNi$_5$ powder was loaded to the sample container until the powder stopped moving even when the sample container was closed and shaken. The thread on the lid was turned so that the height of the container was always 18.5 mm when the case was fully closed (see Fig.\ \ref{Fig:2}). The porosity ($\epsilon$) of the LaNi$_5$ packed bed was estimated from its apparent volume (i.e., $2.12\times 10^4$ mm$^3$), weight, and true density of 7.44 g/cm$^3$ \cite{hahne1998thermal}. As listed in Table \ref{Tab:4}, particle sizes above 150 $\mu$m show relatively small porosity value, which may be caused by the non-uniform size distribution. In contrast, particle sizes below 150 $\mu$m show porosity values close to 0.4, a well-known value for metal-hydrides \cite{pons1991effective, masamune1963thermal}. 

\begin{table}[!t]
	\caption{Porosity ($\epsilon$), residual thermal conductivity ($k_s^0$), and contact factor ($\delta$) of LaNi$_5$ powder packed bed.}
	\label{Tab:4}
	\begin{center}
		\begin{tabular} {c|cccc}
			\hline\hline
			Particle size  & $\epsilon$ & $k_{bulk}$ \cite{hahne1998thermal} & $k_s^0$ & $\delta$ \\
			
			($\mu$m) & & (W/m$\cdot$K) & (W/m$\cdot$K) & \\
			\hline
			below 50      & 0.385 $\pm$ 0.006  & 12.5 & 0.0128 $\pm$ 0.003 & 0.0017\\
			$50\sim 100$       & 0.405 $\pm$ 0.012  & 12.5 & 0.0207 $\pm$ 0.005 & 0.0033\\
			$100\sim150$       & 0.379 $\pm$ 0.009  & 12.5 & 0.0373 $\pm$ 0.005 & 0.0048 \\
			over 150       & 0.317 $\pm$ 0.006  & 12.5 & 0.0363 $\pm$ 0.003 & 0.0043\\
			\hline\hline
		\end{tabular}
	\end{center}
\end{table} 

Theoretical models for the effective thermal conductivity of the packed bed usually assume three parallel heat transfer paths (neglecting radiation and convection in the metal-hydrides due to the small temperature difference with the medium): (1) conduction through contact area among particles; (2) heat transfer through the path of particle-gas film-particle; and (3) conduction through vacant space (i.e., gas conduction). The conductive heat transfer through contact area of particles can be experimentally measured by obtaining the effective thermal conductivity under vacuum condition by eliminating the contribution of gas, and is called \textit{residual thermal conductivity} ($k_s^0$). In our measurements, the pressure of the vacuum chamber gradually reaches 10 mtorr, which can be safely assumed as vacuum condition \cite{pons1991effective}. The pressure inside the sample container becomes equal to this value if the partially cut O-ring is used [see Fig.\ \ref{Fig:2}(b)]. The residual thermal conductivity was measured 3 times for each sample (refer to Table \ref{Tab:4}). Uncertainty of residual thermal conductivity was also estimated by Eq.\ \eqref{Eq:2}. Since $Q_m$ at measurement of residual thermal conductivity is 50\% smaller than that of PEEK, the combined uncertainty is almost four times smaller than that of PEEK. Since three set of measurements were conducted, the sensitivity coefficient of the combined uncertainty was 0.33. Standard deviation of the mean value was under 0.005 W/m$\cdot$K. Therefore, uncertainty of the mean value of residual thermal conductivity were estimated to be less than 0.005 W/m$\cdot$K (i.e., $\sim 25\%$ of the measured $k_s^0$  value). 

\begin{figure}[!t]
    \centering
    \includegraphics[width=0.7\columnwidth]{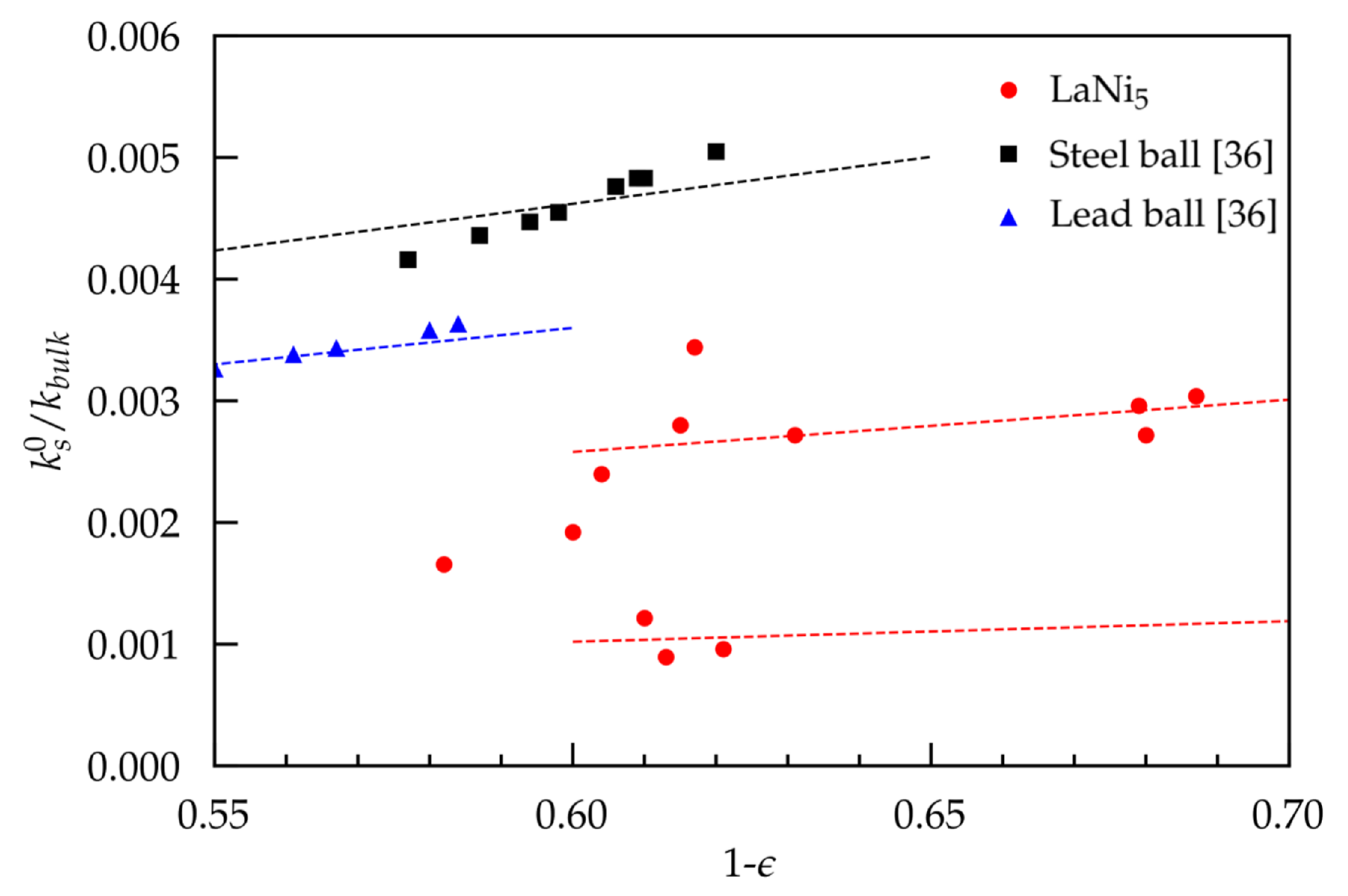}
    \caption{Calculated contact factor of LaNi$_5$ powder packed bed based on the measured residual thermal conductivity ($k^0_s$) and porosity ($\epsilon$). For comparison purpose, $k^0_s$/$k_{bulk}$ of lead balls and steel balls \cite{waddams1944flow} are also plotted.}
    \label{Fig:4}
\end{figure}

Residual thermal conductivity is function of porosity and contact factor that indicates the perfect contact area of the particle. Here, porosity affects the solid fraction of packed bed, and contact factor determines the area that conductive heat transfer occurs. Thus, by using the measured residual thermal conductivity and porosity, the contact factor ($\delta$) of the particles can be obtained from the YK model \cite{yagi1957studies} via:
\begin{equation} \label{Eq:3}
	\delta=\frac{1}{1-\epsilon}\frac{k_s^0}{k_{bulk}}
\end{equation}
where $(1-\epsilon)$ indicates the solid fraction and $k_{bulk}$ is the thermal conductivity of the LaNi$_5$ alloy (12.5 W/m$\cdot$K \cite{hahne1998thermal}). Figure \ref{Fig:4} shows $k^0_s$/$k_{bulk}$ of LaNi$_5$ with respect to solid fraction. For comparison purpose, $k^0_s$/$k_{bulk}$ of lead balls and steel balls \cite{waddams1944flow} are also plotted. The contact factor of LaNi$_5$ is smaller than that of lead and steel ball, which is expected to be caused by the high hardness of LaNi$_5$. Contact factor of the LaNi$_5$ is in the range of $0.004 \sim 0.005$ for the sample of particle size over 100 $\mu$m, and greatly reduce to $0.0017 \sim 0.0033$ when particle size is smaller than 100 $\mu$m. Such trend of the LaNi$_5$ contact factor could be explained based on the surface characteristics and contact area between particles. First, rough surface may increase the contact factor. The contact factor of lead and steel ball shown in Fig.\ \ref{Fig:4} is larger than that of lead and steel shot \cite{masamune1963thermal}, and it was analyzed in Ref.\ \cite{masamune1963thermal} that the lead and steel ball have high contact angle because those particles are sintered to a certain extent, which have rougher surface compare to annealed particles. Thus, in the case of LaNi$_5$, surface is expected to be rougher at particle size over 100 $\mu$m. Second, the larger contact area for sample with particle size over 100 $\mu$m may be due to their non-spherical shape. For verification, the surface characteristics of LaNi$_5$ particles respect to its size were analyzed by SEM.

\begin{figure}[!t]
	\centering
	\includegraphics[width=0.6\columnwidth]{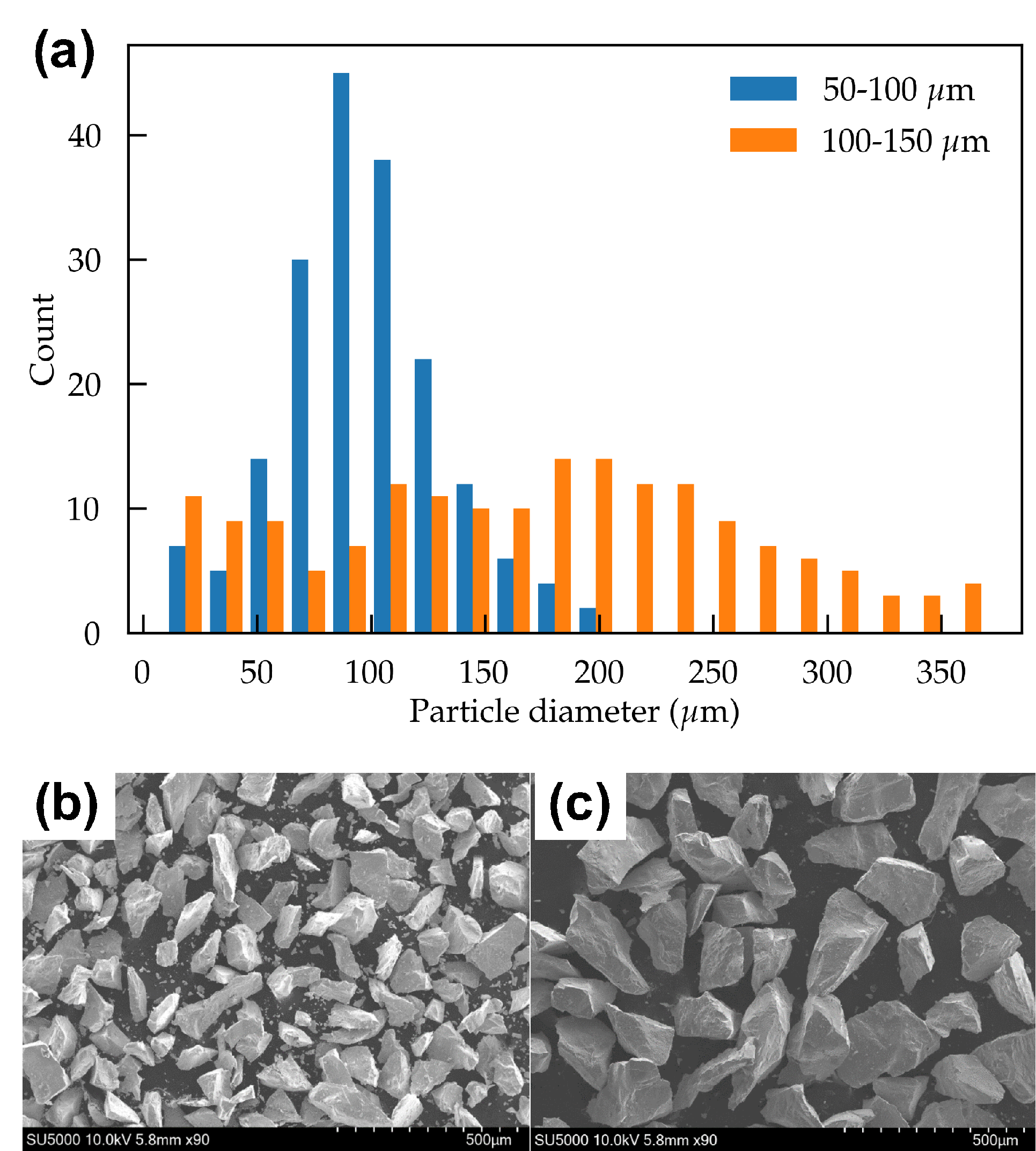}
	\caption{(a) Size distributions of LaNi$_5$ powder sieved through 50-100 $\mu$m and 100-150 $\mu$m meshes. (b) SEM image of the sample sieved with mesh of 50-100 $\mu$m. (c) SEM image of sample sieved with mesh of 100-150 $\mu$m.}
	\label{Fig:5}
\end{figure} 

The size distribution of the LaNi$_5$ powder is shown in Fig.\ \ref{Fig:5}(a). The method described in Ref.\ \cite{rabbani2015comparing} was used for particle diameter calculation. As mentioned earlier, the average particle diameter of the sample was classified using a sieve, and thus, the deviation from the mean value is larger as the aspect ratio of the particle becomes larger. The distribution shown in Fig.\ \ref{Fig:5}(a) implies that the aspect ratio of the sample sieved with $100\sim150$ $\mu$m (Total counts: 173) is higher than the sample sieved with $50\sim100$ $\mu$m (Total counts: 185). This observation is consistent with the fact that the contact factor decreases at particle sizes under 100 $\mu$m. The SEM images shown in Figs.\ \ref{Fig:5}(b) and \ref{Fig:5}(c) also suggest that LaNi$_5$ particles become more regular in shape when they are smaller. As the shape of the particles becomes regular, the proportion of one plane occupying the total surface area of the particles decreases. Thus, area for conductive heat transfer for one contact point decreases. It is also expected that the width of a plane is related to the roughness of the plane. The wider surface is more difficult to maintain flatness than the smaller surface. That is, larger plane of particle size over 100 $\mu$m has more contact point on one plane compare to particle under 100 $\mu$m. 

\begin{figure}[!b]
    \centering
    \includegraphics[width=0.6\columnwidth]{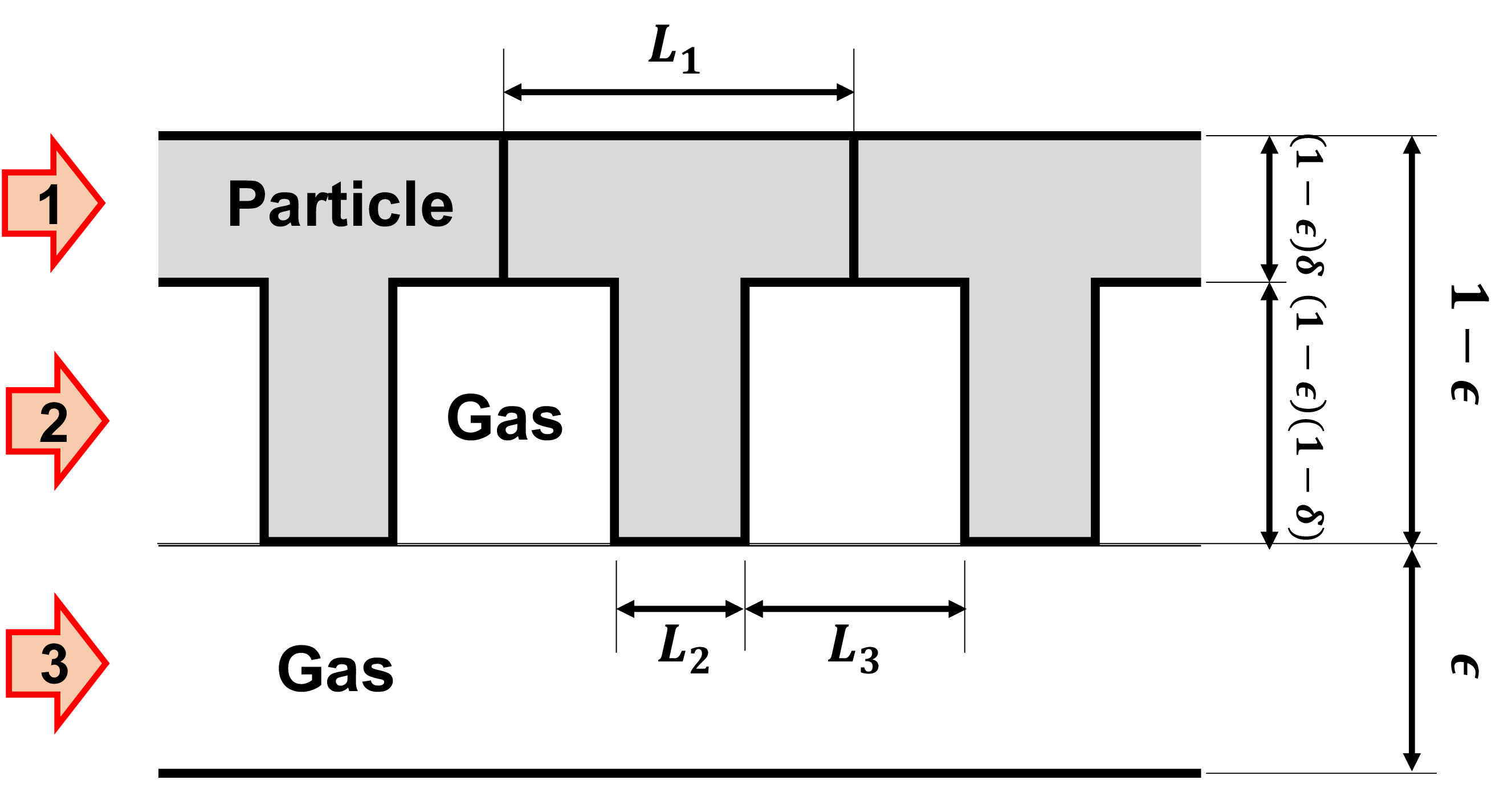}
    \caption{Schematic of the simplified model of heat transfer in packed bed with motionless gas (i.e., YK model). The heat transfer through packed bed is assumed to occur in parallel paths: (1) conduction through contact area among particles; (2) heat transfer through the path of particle-gas film-particle; or (3) conduction through vacant space.}
    \label{Fig:6}
\end{figure}

The different contact factors of LaNi$_5$ compare to those of steel and lead balls suggest the necessity of using the newly characterized thickness of gas film ($\varphi$) for the LaNi$_5$ packed bed. An inaccurate value of $\varphi$ may cause a considerable error in the YK model. Thus, we obtain the value of $\varphi$ for LaNi$_5$ particles using the precisely measured effective thermal conductivity of the LaNi$_5$ packed bed. 

The YK model, combined with the Knudsen effect, has been widely applied for qualitative understanding of the effective thermal conductivity of metal-hydrides \cite{suissa1984experimental,kumar2011measurement,kempf1986measurement}. By simplifying the three heat transfer paths as illustrated in Fig.\ \ref{Fig:6}, the heat transfer through packed bed can be assumed to occur in parallel paths; that is, (1) conduction through contact area among particles (i.e., through $(1-\epsilon) \times \delta$); (2) heat transfer through the path of particle-gas film-particle (i.e., through $(1-\epsilon) \times (1-\delta)$); or (3) conduction through vacant space (i.e, through $\epsilon$). Considering each contribution, total heat transfer $q_{s}$ through the packed bed can be expressed in the YK model as \cite{yagi1957studies}
\begin{equation}
\label{Eq:4}
q_{s}=k_{bulk}(1-\epsilon)\delta\frac{\Delta T}{L_1} +  \frac{(1-\epsilon)(1-\delta)\Delta T}{\frac{L_2}{k_g} + \frac{L_3}{k_{bulk}}} + k_g\epsilon\frac{\Delta T}{L_1}
\end{equation}
where $k_g$ is the gas thermal conductivity, $L_1$ is the characteristic length of the conduction through the contact area, which can generally be expressed as the diameter of the particles (i.e., $D_p$) \cite{yagi1957studies}, and $L_2$ and $L_3$ are the characteristic length of the gas film and the non-contact portion of the particles, respectively, as shown in Fig.\ \ref{Fig:6}. When conduction heat transfer through the contact area is negligible and by noting $L_2=\varphi L_1$, the effective thermal conductivity of packed bed can be further simplified to
\begin{equation}
\label{Eq:5}
\frac{k_s}{k_g}=\frac{1-\epsilon}{\frac{k_g}{k_s}+\varphi}
\end{equation}
where $\varphi$ is the characteristic length of the gas film (i.e., $L_2$/$D_p$). Most studies that have applied YK model to metal-hydrides have used $\varphi$ value of 0.078 \cite{yagi1957studies}, which was empirically estimated for Hydrogen environment with porosity value of $0.3 \sim 0.5$  \cite{suissa1984experimental, kumar2011measurement}. 

\begin{table}[!t]
\caption{Characteristic length of the gas film ($\varphi$) of the LaNi$_5$ packed bed obtained by the measured effective thermal conductivity at Air and Helium condition.}
\label{Tab:5}
\begin{center}
\begin{tabular}{c|cc | cc| c}
\hline
\hline
 Particle size & $k_s$ (in Air) & $\varphi_{Air}$ & $k_s$ (in He) & $\varphi_{\text{He}}$ & $\varphi_{He}$/$\varphi_{Air}$\\
 ($\mu$m) & (W/m$\cdot$K) & & (W/m$\cdot$K) & & \\
 \hline
 below 50 & 0.207 $\pm$ 0.006 & 0.075 & 0.61 $\pm$ 0.20 & 0.141 & 1.87\\
 50 $\sim$ 100 & 0.219 $\pm$ 0.026 & 0.069 & 0.63 $\pm$ 0.20 & 0.130 & 1.88\\
 100 $\sim$ 150 & 0.268 $\pm$ 0.010 & 0.058 & 0.91 $\pm$ 0.33 & 0.090 & 1.56\\
 over 150 & 0.361 $\pm$ 0.014 & 0.048 & 1.65 $\pm$ 0.84 & 0.050 & 1.04\\

\hline\hline
\end{tabular}
\end{center}
\end{table}
\begin{figure}[!b]
    \centering
    \includegraphics[width=0.7\columnwidth]{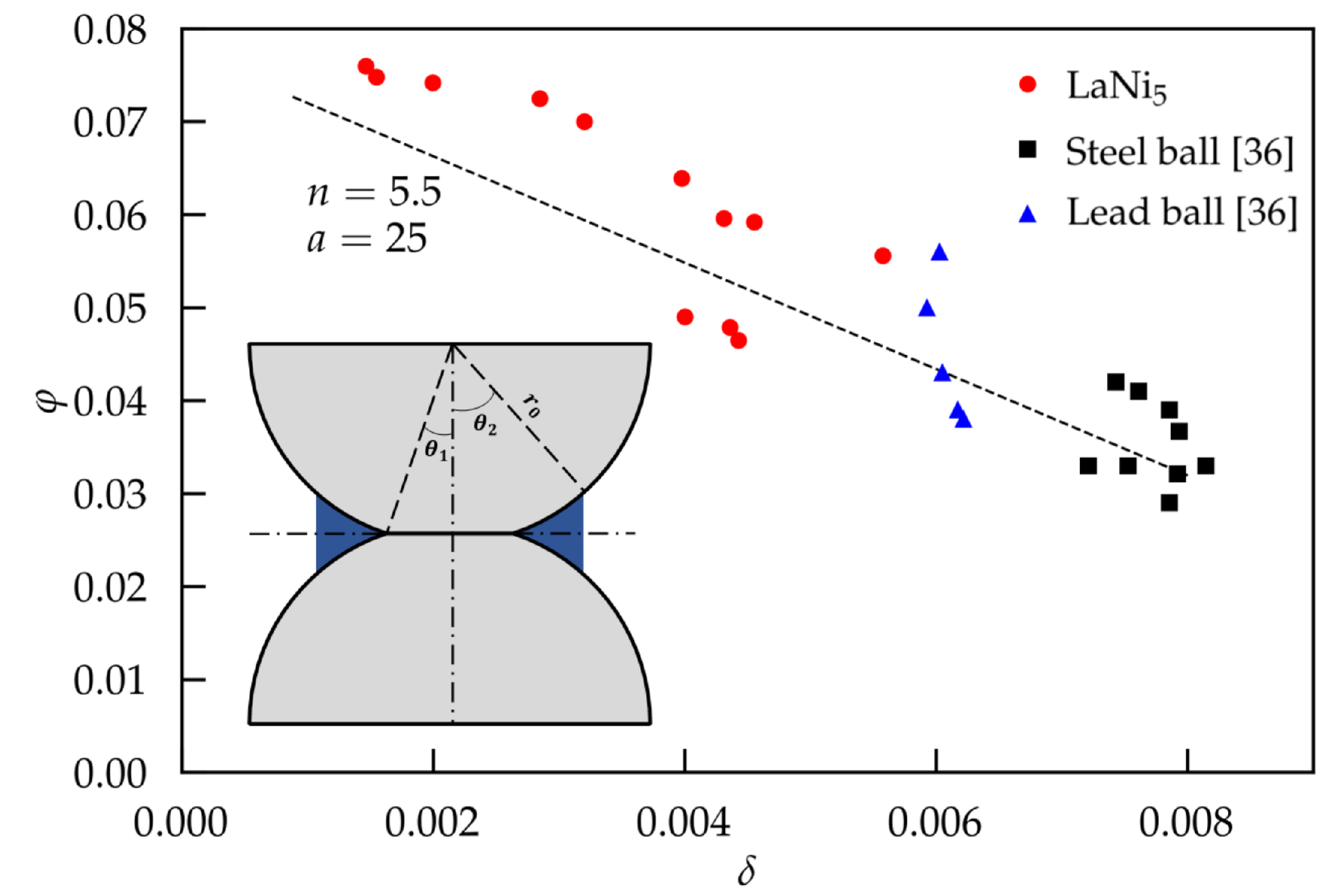}
    \caption{Characteristic length of the gas film ($\varphi$) of LaNi$_5$ powder packed bed at Air condition in atmospheric pressure respect to contact factor. Dashed line is calculation result based on Eqs.\ \eqref{Eq:6}--\eqref{Eq:10}, where $n$ is the average number of contacts on hemispherical area of particle, and $a$ is the sensitivity factor of contact factor. For comparison, $\varphi$ of lead balls and steel balls \cite{waddams1944flow} are also plotted. Inset shows the schematic of simplified model of contact area between particles in packed bed. $\theta_1$ refers contact angle, and region painted in blue represents gas film.}
    \label{Fig:7}
\end{figure}

In this work, $\varphi$ was obtained from the measured effective thermal conductivity of LaNi$_5$ in Air condition at the atmospheric pressure using Eq.\ \eqref{Eq:5}. Air-tight sample container was used to maintain atmospheric pressure in the vacuum chamber. As listed in Table \ref{Tab:5}, the measured effective thermal conductivity was in a range of $0.2 \sim 0.4$ W/m$\cdot$K. If we compare two samples having similar value of contact factor (i.e., particle with $100 \sim 150$ $\mu$m and particle $> 150$ $\mu$m), the characteristic length of the gas film ($\varphi$) could differ by 17\%. Such difference of $\varphi$ at a similar level of contact factor is mainly due to porosity. With decrease of porosity, numbers of contacts between particles increase, and the area of gas film confined by contact points should be also decreased, leading to smaller value of $\varphi$ (quantitative analysis will be given later). On the other hand, if we compare two samples having similar value of porosity (i.e., particle $< 50$ $\mu$m and particle with $100 \sim 150$ $\mu$m), both contact factor and $\varphi$ could be substantially different from each other. Although previous studies also discussed the relationship between porosity and $\varphi$ \cite{masamune1963thermal, yagi1957studies, zehner1970warmeleitfahigkeit, pons1991effective,tsotsas1987thermal}, the contact factor was usually neglected or treated as constant value. However, Table \ref{Tab:4} and Table \ref{Tab:5} clearly indicate that differences in $k_s$ (or $\varphi$) for samples $< 50$ $\mu$m and with $100 \sim 150$ $\mu$m cannot be explained wholly by porosity. This suggests that $\varphi$ does not solely depends on porosity. 

Figure \ref{Fig:7} plots the characteristic length of the gas film ($\varphi$) of LaNi$_5$ with respect to the contact factor ($\delta$). It is obvious from Fig.\ \ref{Fig:7} that there exist approximately linear relationship between $\varphi$ and $\delta$. It is also found that $\varphi$ values of LaNi$_5$ is higher than those of the lead and steel ball with $\epsilon \approx 0.4$ \cite{waddams1944flow}, which is caused by smaller contact factor of LaNi$_5$ particles. Therefore, the contact factor should be taken into consideration when evaluating the characteristic length of the gas film. 

For better understanding, quantitative analysis of the relationship between contact factor ($\delta$) and characteristic length of the gas film ($\varphi$) was also conducted. As shown in the inset of Fig.\ \ref{Fig:7}, $\theta_1$ and $\theta_2$ of the particle determine $\delta$ and $\varphi$. When $n$ is defined as average number of contact points on hemispherical surface of one particle, the proportion of one gas film on the surface of one particle is expressed as \cite{masamune1963thermal}:
\begin{equation}
\label{Eq:6}
\frac{1}{2n}=\frac{1}{4\pi r_{0}^2} \int_{0}^{\theta_{2}}(2\pi r_0 \sin{\theta})r_0 d\theta =\frac{1}{2}(1-\cos{\theta_{2}})
\end{equation}
which leads to 
\begin{equation}
\label{Eq:7}
\theta_2=\cos^{-1} \left(1-\frac{1}{n} \right)
\end{equation}
where $r_0$ is the radius of particle. Because the contact factor determines the proportion of heat transfer in Fig. \ref{Fig:6}, $\theta_1$ and $\theta_2$ can be related to each other via
\begin{equation}
\label{Eq:8}
\frac{\sin^2{\theta_1}}{\sin^2{\theta_2}}=a\delta
\end{equation}
by introducing $a$, the sensitivity factor of $\delta$ on ratio between cross-sectional areas of heat path (1) and (2) [see Fig. \ref{Fig:6}]. For a fixed $\theta_2$, the proportion of gas film at one contact point decreases as contact factor increases (i.e., $\theta_1$ increases). To relate $\varphi$ to $\theta_1$ and $\theta_2$, volume of one gas film ($V_g$) in unit cell was calculated as:
\begin{equation}
\label{Eq:9}
V_g=\int^{\theta_2}_{\theta_1}{r_0 \cos{\theta} d\theta \times 2\pi r_0 \sin{\theta} \times r_0(\cos{\theta_1}-\cos{\theta})}
\end{equation}
and Masamune and Smith \cite{masamune1963thermal} defined $\varphi$ as
\begin{equation}
\label{Eq:10}
\varphi=\frac{L_g}{D_p}=\frac{n V_g}{\pi r_0^2}\frac{1}{2r_0}
\end{equation}
where $L_g$ is thickness of the gas film (i.e., $L_2$). Therefore, through Eq.\ \eqref{Eq:6}--\eqref{Eq:10}, the relationship between $\delta$ and $\varphi$ can be established by taking $n$ and $a$ as fitting parameters. 

As shown in Fig.\ \ref{Fig:7}, calculated $\varphi$ with $n=5.5$ and $a=25$ fits the measured data reasonably well. According to the semi-theoretical expression of $n$ in Ref.\ \cite{kunii1960heat}, $n=1.42$ is for the most open packing (i.e., $\epsilon=0.476$) and $n=6.93$ is for the most dense packing (i.e., $\epsilon=0.260$). Therefore, fitted value of $n=5.5$ suggests the considered LaNi$_5$ power as well as steel and lead balls are more towards the dense packing system. On the other hand, $a=25$ implies that the contact factor affect critically the effective thermal conductivity of the packed bed. The high value of $a$ might be caused by several reasons. One of possible reasons is that when $k_{bulk}$ is much higher than gas thermal conductivity (i.e., 480 times greater than that of air), area near the contact point cannot act as gas film because heat flow through particle will concentrate to contact point due to low thermal resistance. According to the experimental results summarized by Yagi \cite{yagi1957studies}, $\varphi$ increases as the gas thermal conductivity increases. That is, high gas thermal conductivity dissipates the heat flow concentration at the contact point. 

Effect of the gas thermal conductivity was further investigated by measuring the effective thermal conductivity of the LaNi$_5$ packed bed in Helium environment at the atmospheric pressure (also listed in Table \ref{Tab:5}). Considering the instability of Helium molecule (i.e., small molecular weight and diameter), temperature of the heating block was intentionally reduced to 25$\celsius$ to minimize the leaking of Helium through the O-ring. Temperature of the cold plate remain the same (i.e., 10$\celsius$) as before.  The value of $\varphi_{\text{He}}$/$\varphi_{\text{Air}}$ obtained from the measured $k_s$ in He environment was in the range between 1 and 2, which is smaller than the empirical estimation in Ref. \cite{yagi1957studies}. 
\begin{figure}[!b]
    \centering
    \includegraphics[width=0.7\columnwidth]{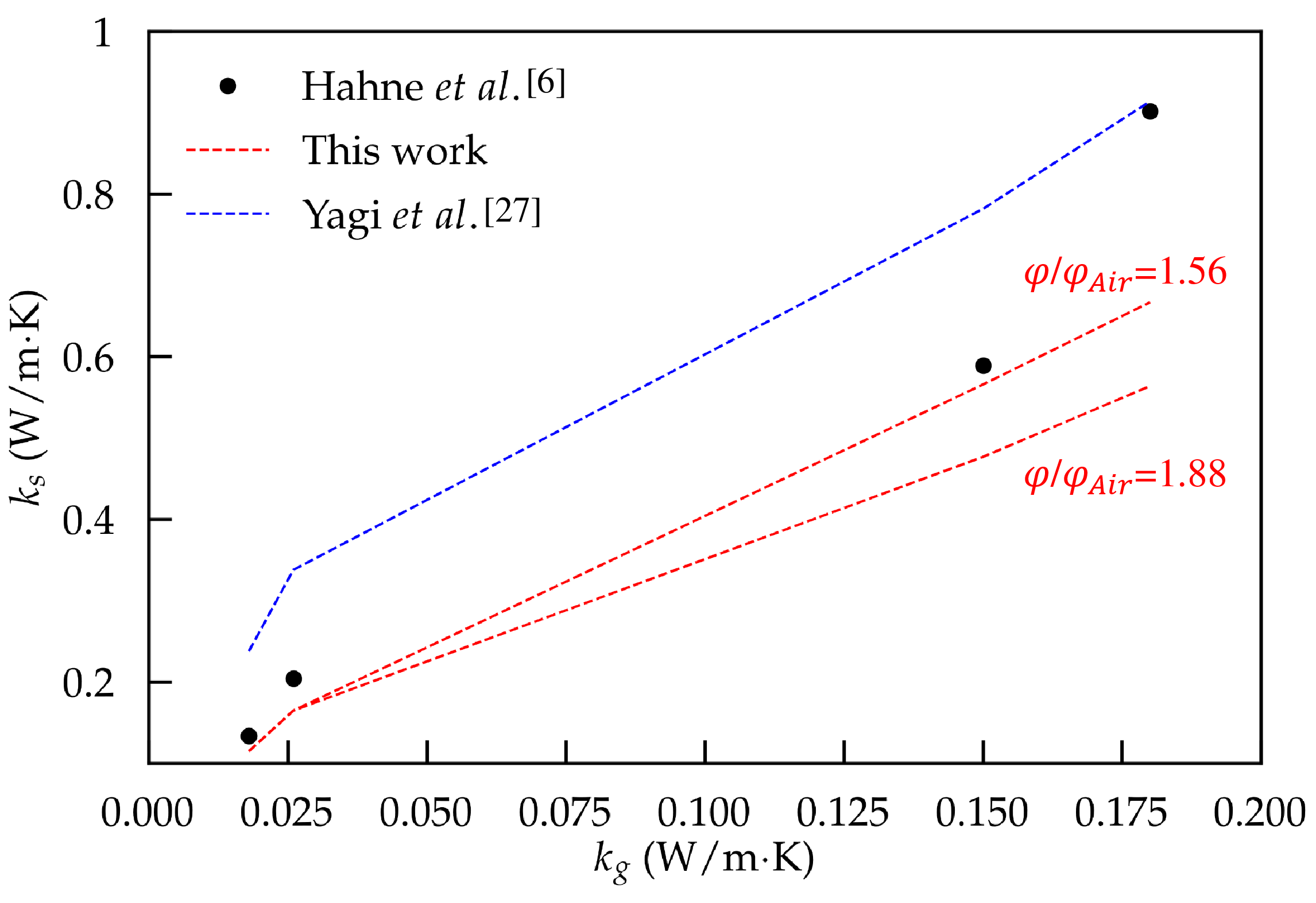}
    \caption{Calculation of the effective thermal conductivity of LaNi$_5$ powder packed bed ($\epsilon$=0.531) for various gas thermal conductivity and comparison with the experimental data \cite{hahne1998thermal}. }
    \label{Fig:8}
\end{figure}

For verification, we calculated $k_s$ of LaNi$_5$ with respect to gas thermal conductivity ($k_g$) by using Eq.\ \eqref{Eq:5} and the obtained $\varphi$ values in the present work. Figure \ref{Fig:8} compares the calculated $k_s$ of LaNi$_5$ with $\varphi$ value either from the empirical estimation \cite{yagi1957studies, suissa1984experimental, kumar2011measurement} or from the present work. Also, experimental data in Ref.\ \cite{hahne1998thermal}, i.e., $k_s$ of LaNi5 at Argon, Air, Helium and Hydrogen environments at the atmospheric pressure, are also plotted for comparison purpose. Here, we used the sample condition, i.e., porosity of 0.531 and $k_s^0=0.005$ W/m$\cdot$K (measured at 0.1 mbar in Hydrogen environment), of Ref.\  \cite{hahne1998thermal} when estimating the contact factor and $\varphi$ (in Air condition) for our model. For Helium (or Hydrogen) environment, we used either $\varphi \approx 1.56 \times \varphi_{Air}$ or $\varphi \approx 1.88 \times \varphi_{Air}$ considering that $\varphi$ can vary with particle sizes (see Table \ref{Tab:5}). It can be seen from Fig.\ \ref{Fig:8} that when $k_g<0.15$ W/m$\cdot$K (i.e., in Argon, Air, or Helium environment), the value of $\varphi$ obtained in the present work results in much better agreement with the measured data than the empirical estimation does.

\section{Conclusion}
In this study, we have analyzed the effective thermal conductivity of LaNi$_5$ powder packed bed with a customized guarded hot- plate (GHP) apparatus. The accuracy of the customized GHP was carefully characterized by uncertainty analysis of the thermal conductivity of PEEK, and showed standard deviation at 4.2\%. We also measured the residual thermal conductivity of LaNi$_5$ powder by using a partially cut O-ring on the sample holder. Based on the measured residual thermal conductivity, effect of particle size on the contact factor of LaNi$_5$ was estimated. The contact factor was found to greatly decrease when the particle size as smaller than 100 $\mu$m. The size distribution of the samples revealed that particle shape becomes more regular as particle size is smaller than 100 $\mu$m, which can eventually decrease the contact factor. By applying the YK model to the effective thermal conductivity of the LaNi$_5$ packed bed, the characteristic length of the gas film ($\varphi$) was obtained to be in the range of 0.04 $\sim$ 0.08 in Air condition at the atmospheric pressure. In addition, by measuring the effective thermal conductivity in Helium environment, the effect of the gas thermal conductivity on $\varphi$ was estimated. It was found that $\varphi$ in Helium environment was 1 $\sim$ 2 times larger than $\varphi$ in Air condition. Our model is capable of predicting the existing experimental data excellently for wide range of $k_g$ (i.e., $0.018 \sim  0.15$ W/m$\cdot$K). The results obtained in this study will be helpful for designing an efficient thermal management for hydrogen storage systems using metal-hydride.



\section*{Acknowledgments}

This research was supported by the International Energy Joint R\&D Program of the Korea Institute of Energy Technology Evaluation and Planning (KETEP), granted financial resources by the Ministry of Trade, Industry \& Energy, Republic of Korea (No.\ 20188520000570), as well as by the Agency for Defense Development in South Korea (ADD).

\section*{References}
\bibliographystyle{elsarticle-num} 
\bibliography{Refs}





\end{document}